\documentclass[aps,pra,floatfix,twocolumn,superscriptaddress,showpacs,10pt,]{revtex4-1}
\usepackage{amssymb, amsmath,color,mciteplus,graphicx,subfigure}
\usepackage{calc,epsfig,epstopdf,color,mciteplus,bm,mathrsfs}
\usepackage{times}
\usepackage{float}
\usepackage{lipsum}
\usepackage{booktabs}

\begin{document}

\title{Nonadiabatic geometric quantum gates that are insensitive to qubit-frequency drifts}

\author{Jian Zhou}
\affiliation{School of Electrical and Opto-Electronic Engineering, West Anhui University, Lu'an 237012, China}
\affiliation{Department of Electronic Communication Engineering, Anhui Xinhua University, Hefei, 230088, China}

\author{Sai Li}
\affiliation{Guangdong Provincial Key Laboratory of Quantum Engineering and Quantum Materials, and School of Physics\\ and Telecommunication Engineering, South China Normal University, Guangzhou 510006, China}

\author{Guo-Zhu Pan}
\affiliation{School of Electrical and Opto-Electronic Engineering, West Anhui University, Lu'an 237012, China}

\author{Gang Zhang}\email{zhanggang@wxc.edu.cn}
\affiliation{School of Electrical and Opto-Electronic Engineering, West Anhui University, Lu'an 237012, China}

\author{Tao Chen}\email{chentamail@163.com}
\affiliation{Guangdong Provincial Key Laboratory of Quantum Engineering and Quantum Materials, and School of Physics\\ and Telecommunication Engineering, South China Normal University, Guangzhou 510006, China}

\author{Zheng-Yuan Xue}\email{zyxue83@163.com}
\affiliation{Guangdong Provincial Key Laboratory of Quantum Engineering and Quantum Materials, and School of Physics\\ and Telecommunication Engineering, South China Normal University, Guangzhou 510006, China}

\affiliation{ Guangdong-Hong Kong Joint Laboratory of Quantum Matter, and Frontier Research Institute for Physics,\\ South China Normal University, Guangzhou  510006, China}

\date{\today}

\begin{abstract}
  Quantum manipulation based on geometric phases provides a promising way towards  robust quantum gates. However, in the current implementation of nonadiabatic geometric phases, operational and/or random errors tend to destruct the conditions that induce geometric phases, thereby smearing their noise-resilient feature. In a recent experiment [Y. Xu \emph{et al}., Phys. Rev. Lett. \textbf{124}, 230503 (2020)], high-fidelity universal geometric quantum gates have been implemented in a superconducting circuit, which are robust to different types of errors under different configurations of the geometric evolution paths. Here, we apply the path-design strategy to explain in detail why both configurations can realize universal quantum gates in a single-loop way. Meanwhile, we purposefully induce our geometric manipulation by selecting the path configuration that is robust against the qubit-frequency-drift induced error, which is the dominant error source on realistic superconducting circuits and has not been deliberately addressed. Moreover, our proposal can further integrate with the composite scheme to enhance the gate robustness, which is verified by numerical simulations. Therefore, our scheme provides a promising way towards practical realization of high-fidelity and robust nonadiabatic geometric quantum gates.
 \end{abstract}

\maketitle

\section{introduction}

Geometric phases \cite{Berry1984a, Wilczek1984b, Aharonov1987} unveil important geometric structures of the evolution quantum states during quantum dynamics. Different from the dynamical phases, geometric phases are determined by the global property of the evolution paths, so that they are largely insensitive to many local noises \cite{Zhu2005, Solinas2012, Johansson2012} and have found many important applications  nowadays \cite{Xiao2010}. In contrast to the adiabatic cases \cite{PZ1999, JP1999, LM2001, PZ2005, IK2011, VVA2016}, the built-in noise resilience of the nonadiabatic geometric phases provide a more practical way to implement quantum computation and has been recently proposed based on both Abelian \cite{Wang2001, Zhu2002, Zhu2003a, Tian2004, Zhao2017, Chen2018a} and non-Abelian geometric phases \cite{Sjoqvist2012, Xu2012, Zhang2014d, VA14a, VA14b, Xu2015, zhao16, zhao17, Xu2017, Xue2017, ZJ18, Hong2018, ChenAn, Liu19, Ji2019, Li2020, zhao2020, ChenTOC}. Therefore, many renewed efforts have recently been given to their experimental demonstration in various quantum systems \cite{Pechal2012, Abdumalikov2013, Feng2013, Zu2014, Li2017a, Zhou2017c, Xu2018, Yan2019, zhu2019, Xu2020}.

Meanwhile, the superconducting quantum circuits system \cite{Koch2007, Clarke2008, You2011a, Devoret2013c, Xiang2013a} is a promising candidate for quantum information processing, due to its distinct merits in scalability and stability. But, the coupling strength  in the simplest capacitive coupled superconducting qubits is hard to be tuned. Fortunately,  parametrically tunable coupling \cite{Niskanen2007, McKay2016, Lu2017, Naik2017, Reagor2018a, Li2018, Caldwell2018, ExpYuY} between two superconducting qubits with different frequencies can be obtained by adding an ac driving on one of the qubits to periodically modulate its transition frequency. Thus, with this technique, one can realize controllable interaction between adjacent qubits in the simplest circuits when scaling up the qubit lattice.

Generally, the implementation of nonadiabatic geometric quantum computation (NGQC), particularly nontrivial two-qubit gates, is very difficult due to the control imperfections and randomized qubit-frequency-drift-induced error, which respectively represent the $\sigma_x$ and $\sigma_z$ errors in superconducting quantum circuits system and will inevitably induce errors in the target quantum gates. Moreover, the need for complex control of multilevel quantum systems and the intrinsic leakage of the quantum information are also nonnegligible error sources.  Remarkably, theoretic NGQC schemes \cite{Tian2004, Zhao2017, Chen2018a} have been proposed based on only effective two-level systems and two-body interactions, which can effectively suppress the $\sigma_x$ error by using the composite strategy \cite{Chen2018a}. Experimentally, a recent elementary demonstration \cite{Xu2020} in a superconducting circuit has verified that geometric gates can be robust against two different types of quantum errors with two different configuration settings. The enhancement of the robustness of geometric gates against the $\sigma_x$ error has also been experimentally verified using the pulse optimization \cite{Yan2019} and composite \cite{zhu2019} schemes. But, for superconducting quantum circuits, once the pulse envelope is calibrated, the $\sigma_x$ error can be stabilized within a deviation of a percent of its typical Rabi frequency. However, as to the $\sigma_z$ error, it can be up to a few hundreds of kilohertz, i.e., several percentages of the typical Rabi frequency, thus being the dominant error source. Note that the frequency drift of a superconducting qubit is usually of static nature; i.e., it is a constant during a gate-operation time but may be randomly changed among different gates.

\begin{figure*}[tbp]
  \centering
  \includegraphics[width=0.75\linewidth]{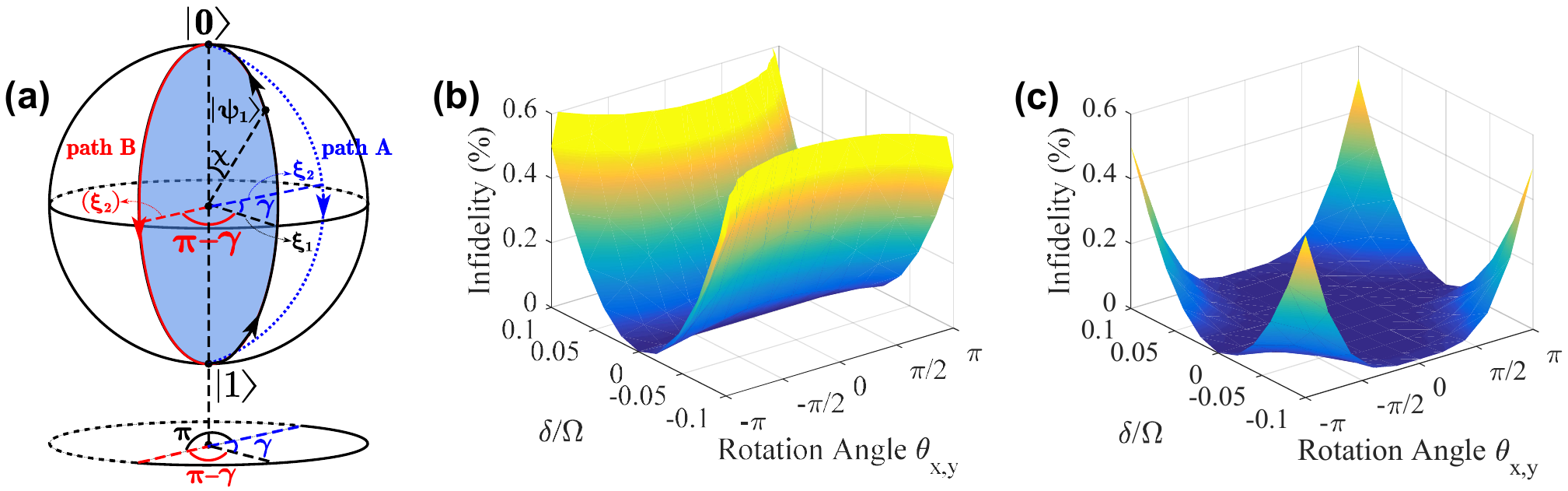}
  \caption{(a) Illustration of geometric quantum gates induced from different evolution paths, i.e., configurations A and B. Noise-resilient feature against the qubit-frequency drift $\sigma_z$ error (b) in configuration A and (c) in configuration B.} \label{PathAB}
\end{figure*}

Here, we first apply the path-design strategy to explain why the experiment in Ref. \cite{Xu2020}  has two configurations that can both realize universal geometric gates in a single-loop way. Meanwhile, we can select the most robust path configuration, and then show that the composite scheme can be introduced to implement geometric quantum gates that are insensitive to the qubit-frequency-drift induced $\sigma_z$ error. In addition, we further exhibit the gate performance on a scalable two-dimensional (2D) square lattice with capacitive coupled superconducting qubits, where we only use an effective two-level system and two-body interaction, and the leakage of quantum information can also be effectively suppressed. Finally, our numerical simulations verify the improvement of our geometric gate robustness while maintaining high fidelity, and gate robustness is even superior to conventional dynamical gates \cite{DYG2014}. Therefore, our scheme provides a promising method  to achieve high-fidelity geometric manipulation for robust and scalable solid-state quantum computation.

\section{Geometric Path Design}

Generally, the dynamics of a quantum system are captured by the evolution of its eigenstates. Here, we aim to elaborate how to design a suitable evolution path for a target geometric gate. We first proceed to a general two-level system in an ideal situation, under the interaction framework, assuming $\hbar=1$ hereafter; the reduced Hamiltonian in the computational bases $|0\rangle$ and $|1\rangle$ is
\begin{eqnarray}
\label{h0}
\mathcal{H}(t)=\frac {1} {2}
\left(
\begin{array}{cccc}
 -\Delta(t)             & \Omega(t) e^{-\textrm{i}\phi(t)} \\
 \Omega(t) e^{\textrm{i}\phi(t)} & \Delta(t)
\end{array}
\right),
\end{eqnarray}
where $\Omega(t)$ and $\phi(t)$ are the amplitude and phase of the driving microwave field, respectively, and $\Delta(t)$ is the time-dependent detuning between the qubit transition frequency and the frequency of the microwave field.

To get the target geometric gates and visualize their evolution details under the driven Hamiltonian $\mathcal{H}(t)$, we choose a pair of orthogonal dressed-state bases, i.e.,
\begin{eqnarray}
|\psi_1(t)\rangle&=&\cos{\frac {\chi(t)} {2}}|0\rangle+\sin{\frac {\chi(t)} {2}}e^{\textrm{i}\xi(t)}|1\rangle,\notag\\
|\psi_2(t)\rangle&=&\sin{\frac {\chi(t)} {2}}e^{-\textrm{i}\xi(t)}|0\rangle-\cos{\frac {\chi(t)} {2}}|1\rangle,
\end{eqnarray}
where $\chi(t)$ and $\xi(t)$ are time-dependent parameters, representing the polar and azimuthal angles of these evolution states in a Bloch sphere, respectively, as shown in Fig. \ref{PathAB}(a). By letting the dressed states satisfy the von Neumann equation of
\begin{eqnarray}
\label{vN}
\frac{\partial} {\partial t}\left(|\psi_{j}(t)\rangle \langle \psi_j(t)|\right)=-\textrm{i}[\mathcal{H}(t), |\psi_j(t)\rangle \langle \psi_j(t)|],
\end{eqnarray}
with $j=1,2$, we can obtain
\begin{eqnarray}
\label{conditions}
\dot{\xi}(t)&=&-\Delta(t)-\Omega(t)\cot\chi(t)\cos[\phi(t)-\xi(t)], \notag \\
\dot{\chi}(t)&=&\Omega(t)\sin[\phi(t)-\xi(t)],
\end{eqnarray}
which indicate that, arbitrary time-dependent variations of the required evolution parameters $\chi(t)$ and $\xi(t)$ can all be realized through the changing of the Hamiltonian parameters $\{\Omega(t),\Delta(t),\phi(t)\}$. Therefore, we can drive the evolution states $|\psi_{1,2}(t)\rangle$ to follow different evolution paths. After a period of cyclical evolution, at the final time $\tau$, these two dressed states will be $|\psi_j(\tau)\rangle=e^{ \textrm{i}\gamma_j}|\psi_j(0)\rangle$, and the corresponding evolution operator is
\begin{eqnarray}
\label{EqU}
U(\tau)=\sum_{j=1,2}e^{ \textrm{i}\gamma_j}|\psi_j(0)\rangle\langle \psi_j(0)|,
\end{eqnarray}
with $\gamma_1=-\gamma_2=\gamma_d+\gamma_g$ being a total phase, and the dynamical phase part is
\begin{eqnarray}
\label{Phased}
\gamma_d&=& -\int^\tau_0\langle \psi_1(t)| \mathcal{H}(t)|\psi_1(t)\rangle \textrm{d}t \notag \\
& =&\frac {1} {2}\int^\tau_0 \frac{\dot{\xi}(t)\sin^2\chi(t)+\Delta(t)}{\cos\chi(t)} \textrm{d}t.
\end{eqnarray}
The remaining part,
\begin{eqnarray}
\label{Phaseg}
\gamma_g& =&\textrm{i}\int^\tau_0\langle \psi_1(t)|\frac {\partial} {\partial t}|\psi_1(t)\rangle \textrm{d}t\notag \\
& =&-\int^\tau_0\frac{\dot{\xi}(t)}{2}[1-\cos\chi(t)]\textrm{d}t,
\end{eqnarray}
is the geometric phase, which just represents half of the solid angle enclosed by the evolution path in the Bloch sphere. In the following, we utilize the path-design strategy to reverify how to realize an universal geometric evolution in  different single-loop ways \cite{Xu2020}. Moreover, we show how to realize a noise-resilient geometric evolution based on a composite-loop method in the presence of the dominant error source, i.e., the qubit-frequency-drift-induced $\sigma_z$  error.

\begin{figure*}[tbp]
  \centering
  \includegraphics[width=0.85\linewidth]{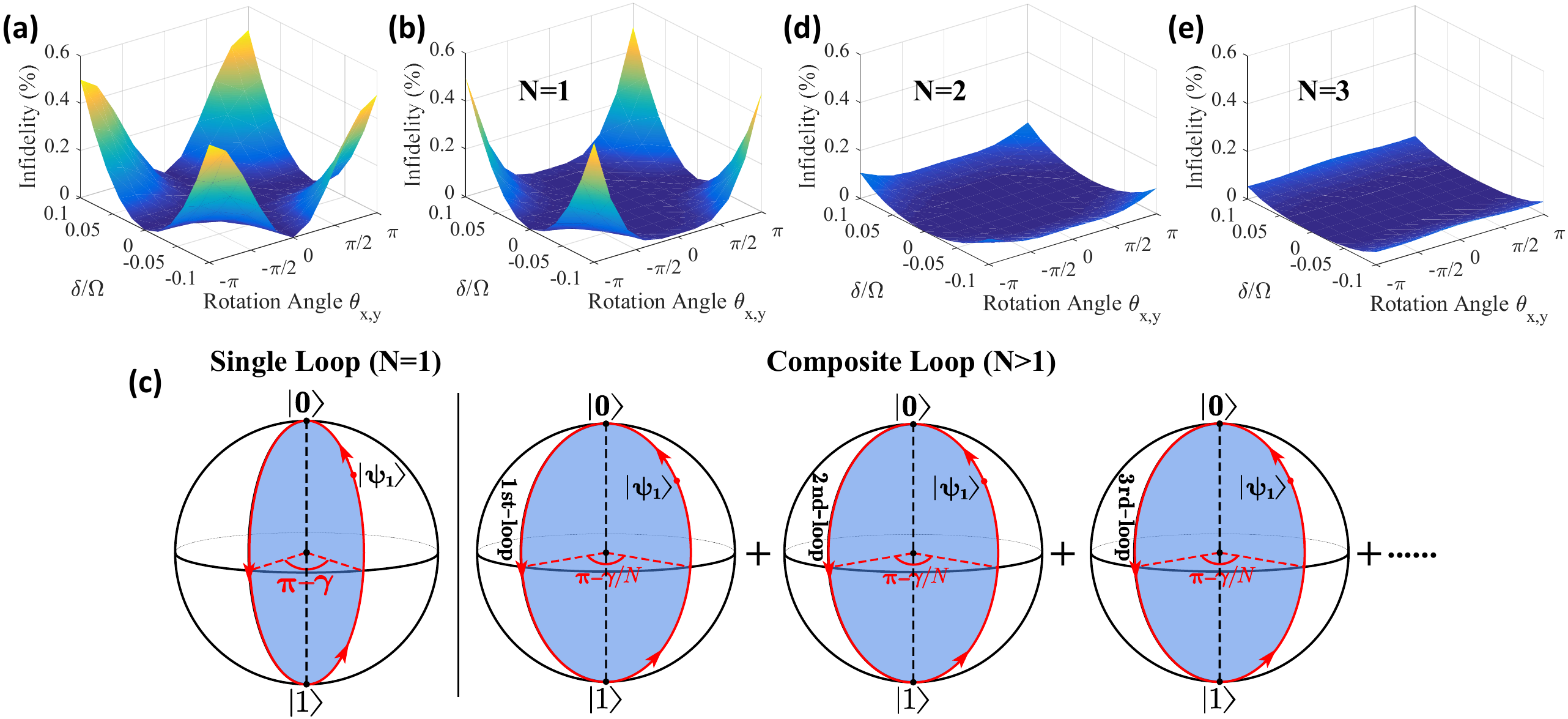}
  \caption{The gate infidelities versus the qubit-frequency-drift-induced $\sigma_z$ error for (a) conventional dynamical gates, (b) single-loop $(N\!=\!1)$ and (d,e) composite $(N\!>\!1)$ geometric gates. (c) Single-loop and composite-loop geometric evolution trajectories in the Bloch sphere.} \label{composite1}
\end{figure*}

First, to realize a pure geometric evolution in the resonant case, i.e., $\Delta(t)=0$, we set $\xi(t)=\phi(t)\pm\pi/2$ to completely eliminate the dynamical phase, i.e., $\gamma_d=0$. Thus, the constraints for the dressed-state parameters in Eq. (\ref{conditions}) reduce to $\dot{\xi}(t)=0$ and $\chi(t)=\chi(0)\mp\int^\tau_0 \Omega(t)\textrm{d}t$. In this case, a cyclical evolution path is designed as follows
\begin{eqnarray}
\label{path}
\chi(0)\!=\!\chi \longrightarrow &\chi(\tau_1)\!=\!0 &\longrightarrow \ \ \chi(\tau_2)\!=\!\pi \ \ \longrightarrow  \chi(\tau)\!=\!\chi, \notag\\
\xi(0)\!=\!\xi_1\!\! \longrightarrow &
\!\begin{array}{cccc}
 \xi(\tau_1\!-\!\varepsilon)\!=\!\xi_1              \\
 \xi(\tau_1\!+\!\varepsilon)\!=\!\xi_2
\end{array}\!  &\longrightarrow \!\!\begin{array}{cccc}
 \xi(\tau_2\!-\!\varepsilon)\!=\!\xi_2              \\
 \xi(\tau_2\!+\!\varepsilon)\!=\!\xi_1
\end{array}\!\!\! \longrightarrow \! \xi(\tau)\!=\!\xi_1, \notag \\
\end{eqnarray}
where, in detail, the polar angle changes from the initial value $\chi$ to the north pole, and then back to $\chi$ from the south pole. Meanwhile, the azimuthal angle remains unchanged on the longitude, only changing from $\xi_1$ to $\xi_2$ and from $\xi_2$ back to $\xi_1$ at the north pole and south pole, i.e., the intermediate time points $\tau_1\mp\varepsilon$ and $\tau_2\mp\varepsilon$, respectively, in which $\varepsilon\!\ll\!0$ is the instantaneous jump time. The evolution details can also be visualized in a Bloch sphere, as shown in Fig. \ref{PathAB}(a), which validates that half of the solid angle enclosed by this single-loop path is exactly equal to the geometric phase $\gamma_g=\xi_2-\xi_1$ \cite{Zhao2017,Chen2018a}. In addition, the selection of the dressed-state parameters $\chi$, $\xi_1$, and $\xi_2$ depends on the target gate type. Thus, the resulting geometric evolution operator, at the final time $\tau$, is
\begin{eqnarray}
\label{Utau}
U(\gamma_g)&=& U(\tau,\tau_2)U(\tau_2,\tau_1)U(\tau_1,0)\nonumber \\
&=&\cos \gamma_g + \textrm{i}\sin\gamma_g \left(\begin{array}{ccc}
\cos\chi & \sin\chi e^{-\textrm{i}\xi_1} \\
\sin\chi e^{\textrm{i}\xi_1} & -\cos\chi
\end{array}\right).
\end{eqnarray}
Moreover, we find that there are two different evolution paths, as shown in Fig. \ref{PathAB}(a), namely $\xi_2-\xi_1=\gamma$ and $\gamma-\pi$, which can both be used to realize the universal geometric operations. That just reverifies and intuitively explains why two different geometric evolution paths in Ref. \cite{Xu2020} can be selected, i.e., ``configuration A" and ``configuration B" there.

Based on the restrictions of dressed-state parameters in Eq. (\ref{path}), we next return the settings of the corresponding Hamiltonian parameters. In configuration A, the entire orange-slice-shaped evolution can be realized with a three-component microwave drive, with the driving amplitude $\Omega(t)$ and phase $\phi(t)$ of the three time intervals $0 \rightarrow \tau_1$, $\tau_1 \rightarrow \tau_2$, and $\tau_2 \rightarrow \tau$, satisfying
\begin{eqnarray}\label{divide+}\begin{cases}
\int_0^{\tau_1}\Omega(t)dt=\chi,  \quad \phi(t)=\xi_1- \pi/ 2,  \quad t\in[0, \tau_1], \\ \\
\int_{\tau_1}^{\tau_2}\Omega(t) dt=\pi,  \quad \phi(t)=\xi_1+\gamma+\pi/ 2, t\in[\tau_1, \tau_2], \\ \\
\int_{\tau_2}^{\tau}\Omega(t) dt=\pi-\chi, \ \ \phi(t)=\xi_1-\pi/ 2, \quad t\in[\tau_2, \tau].
\end{cases}\end{eqnarray}
And in configuration B, the geometric evolution is realized by setting the phase $\xi_1+\gamma-\pi/ 2$ at the $t\in[\tau_1, \tau_2]$ interval in Eq. (\ref{divide+}), while the resulting geometric operators are the same as in Eq. (\ref{Utau}). Then, arbitrary single-qubit nonadiabatic geometric gates can be realized in a single-loop way with two different evolution trajectories.

However, as experimentally demonstrated in Ref. \cite{Xu2020}, the gate performance is usually influenced by the chosen evolution path. Therefore, it is crucial to find a path that is most resistant to the dominant errors in these two alternative geometric paths A and B. In current quantum experiment platforms, the qubit-frequency drift, in the form of $\delta|1\rangle\langle 1|$ with $\delta=\delta_0\Omega$ being the drift quantity, becomes one of the main error sources, where $\Omega(t)=\Omega$ is set to be a square pulse for simplicity. In the presence of the qubit-frequency drift, the Hamiltonian in Eq. (\ref{h0}) turns to $\mathcal{H}^{\delta}(t)=\mathcal{H}(t)+\delta|1\rangle\langle 1|$. To fully estimate the sensitivity of elementary gates to qubit-frequency drift, under the two alternative geometric configurations A and B, we calculate the fidelity of these gates by using the formula \cite{Wang2009b}
\begin{eqnarray}
\label{simulate1}
F_{U_\delta}=  {\text{Tr} (U^\dag U_\delta)}/{\text{Tr}(U^\dag U)},
\end{eqnarray}
where $U$ and $U_\delta$ represent a target gate and the gate affected by the qubit-frequency drift, respectively. Here, we choose the X- and Y-axis rotation operations $\textrm{R}_{x,y}(\theta_{x,y})$ with different rotation angles $\theta_{x,y}$ for our demonstration purpose, which are two types of noncommutating gates that are a universal set for single-qubit quantum gates. For our geometric quantum gates, $\textrm{R}_{x,y}(\theta_{x,y})$ can be realized by setting $\xi_1$ to $\pi$ and $-\pi/2$ when $\chi=\pi/2$ and $\gamma=\theta_{x,y}/2$. By numerically simulating the infidelity $1-F_{U_\delta}$ of the geometric gates in the presence of the qubit-frequency drift, as shown in Figs. \ref{PathAB}(b) and \ref{PathAB}(c), one can clearly see that the well-designed evolution path in configuration B has distinct advantages over  configuration A in terms of robustness against the qubit-frequency-drift induced $\sigma_z$ error. To sum up, considering that if the dominant error-source in an experimental setup is the $\sigma_z$ error, we purposefully select the path in configuration B to implement universal geometric gates.

\section{Noise-resilient Geometric Gates}

\begin{figure}[tbp]
 \centering
 \includegraphics[width=0.9\columnwidth]{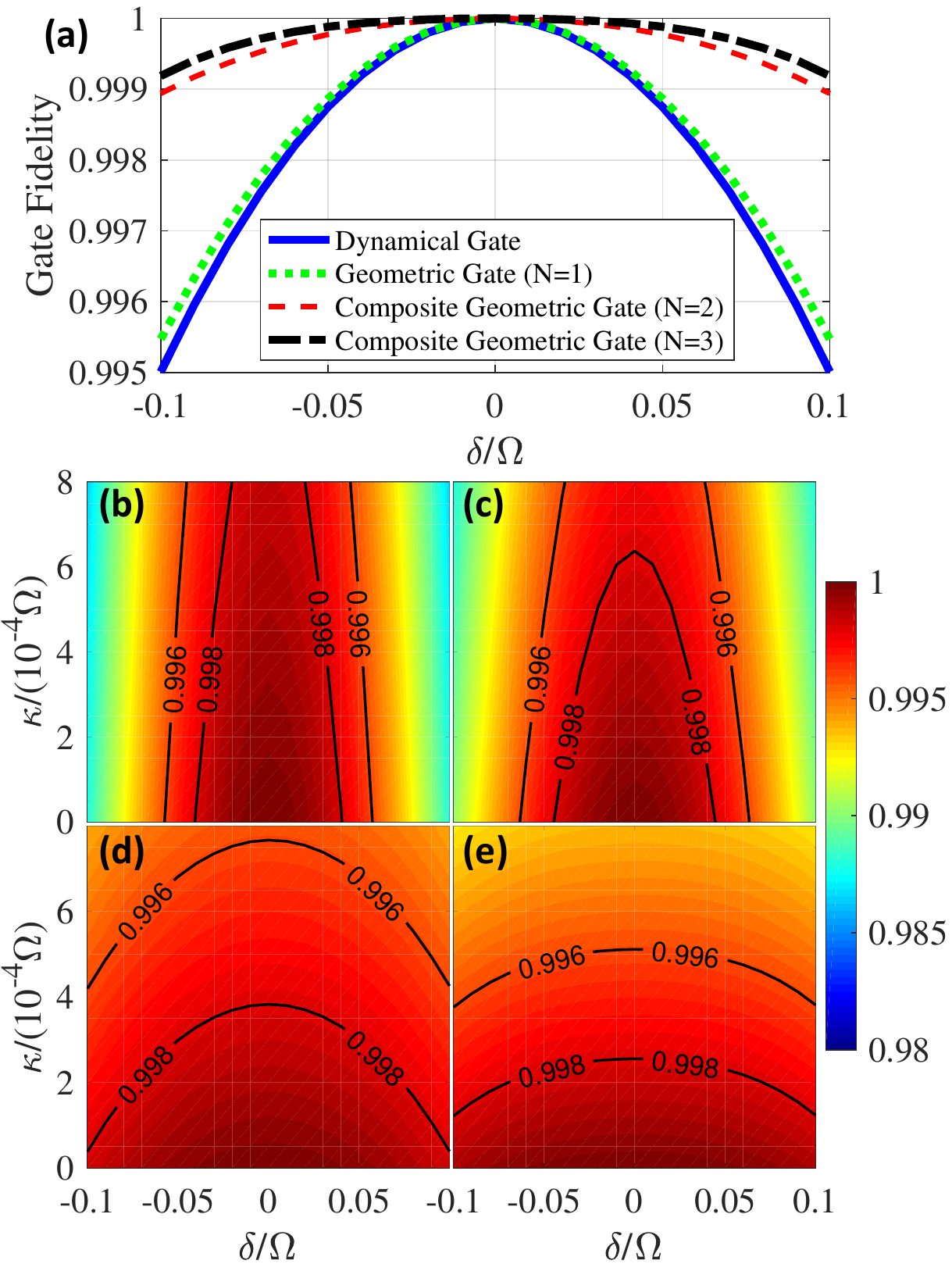}
  \caption{(a) NOT gate fidelities as a function of the qubit-frequency drift without decoherence. Gate fidelities under the qubit-frequency drift and decoherence for (b) the conventional dynamical NOT gate, (c) single-loop $(N=1)$ and (d,e) composite $(N=2,3)$ geometric NOT gates, respectively.} \label{composite2}
\end{figure}

Generally, the existence of quantum errors will inevitably cause the destruction of geometric conditions in Eq. (\ref{Phaseg}), so our handpicked geometric gates based on configuration B do not have an obvious advantage over the dynamical counterparts \cite{DYG2014} (see Appendix A for details), as shown in Figs. \ref{composite1}(a) and \ref{composite1}(b). To further enhance geometric gate robustness against the qubit-frequency-drift-induced $\sigma_z$ error on a whole, we here propose a composite scheme for the implementation of noise-resilient geometric gates. We take $U_c(\gamma_g)$ with $\gamma_g\!=\!\gamma^c_g\!=\!\gamma/N - \pi$ in Eq. (\ref{Utau}) as the elementary gate and sequentially apply the elementary gate ${N}$ times with ${N} > 1$; the details of the evolution path are shown in Fig. \ref{composite1}(c). Then, the following composite geometric gate (ignore a global phase),
\begin{eqnarray}\label{compU}
U_c({N}\gamma^c_g) = [U_c(\gamma^c_g)]^N =\sum_{j=1,2}e^{ \textrm{i}\gamma}|\psi_j(0)\rangle\langle \psi_j(0)|,
\end{eqnarray}
can be obtained. The gate infidelities of composite geometric gates with $N=2, 3$ as a function of the qubit-frequency drift are plotted in Figs. \ref{composite1}(d) and \ref{composite1}(e) under the same driving strength. We find that composite geometric gates greatly improve the robustness against qubit-frequency-drift-induced error and exhibit obvious advantages over the conventional dynamical gates as well as single-loop ($N$=1) geometric gates. Note that a similar discussion is available for the case where the qubit-frequency drift is not static but fluctuates on a time-scale shorter than the gate-operation time (for details, see Appendix B); our composite geometric gates can still exhibit obvious advantages of gate robustness.

\begin{table}
\caption{Gate-performance comparison of the NOT gates from different implementations.}
\label{tab}
\begin{tabular}{lll}
\hline\noalign{\smallskip}
\quad \quad \quad \quad Get type & \quad  Gate fidelity \quad &  Gate time  \\
\noalign{\smallskip}\hline\noalign{\smallskip}
  \quad \quad \quad Dynamical gate  &  \quad $1-(\pi/6){\delta_0}^2$    & \quad \ \ $\pi/\Omega$ \\

  \quad \quad Geometric Gate (N=1)  &  \quad $1-(\pi/7){\delta_0}^2$ & \quad $2\pi/\Omega$ \\

  Geometric Composite Gate (N=2)    &  \quad $1-O({\delta_0}^3)$  & \quad $4\pi/\Omega$ \\

  Geometric Composite Gate (N=3)    &  \quad $1-O({\delta_0}^3)$   & \quad $6\pi/\Omega$ \\

\noalign{\smallskip}\hline
\end{tabular}\\
\end{table}

Moreover, the composite scheme can enhance gate robustness more strongly with larger $N$, as shown in Fig. \ref{composite2}(a), by taking the NOT gate without decoherence as a typical example. The theoretical results using Eq. (\ref{simulate1}), listed in Table \ref{tab}, also agree with the numerical results. However, as listed in Table \ref{tab}, the composite geometric gate requires a multifold increase in the evolution path or, equivalently, the gate-time cost; thus we need to balance the influence from  decoherence and qubit-frequency drift.
Considering that the decoherence process is unavoidable, we further take the decoherence effect into consideration and simulate the evolution process numerically by the master equation \cite{Motzoi2009} of
\begin{eqnarray}
\label{simulate2}
\dot\rho=-i[\mathcal{H}^{\delta}(t), \rho]+\left[{\kappa_-}\mathcal{L}(|0\rangle\langle 1|)+{\kappa_z}\mathcal{L}(|1\rangle\langle 1|)\right]/2,\quad
\end{eqnarray}
where $\rho$ is the reduced density matrix of the considered quantum system, and $\mathcal{L}(\mathcal{A}) = 2\mathcal{A}\rho\mathcal{A^\dag}-\mathcal{A^\dag}\mathcal{A}\rho-\rho\mathcal{A^\dag}\mathcal{A}$ is the Lindblad operator for operator $\mathcal{A}$ with $\kappa_-$ and $\kappa_z$ being the decay and dephasing rates of the qubit, respectively. For the general initial state of a single qubit as $|\psi_1\rangle=\cos\theta_1|0\rangle+\sin\theta_1|1\rangle$ with $|\psi_{f}\rangle=\sin\theta_1|0\rangle+\cos\theta_1|1\rangle$ being the ideal final state, we define single-qubit gate fidelity as $F_{_{\textrm{NOT}}}^G=\frac {1} {2\pi}\int_0^{2\pi} \langle \psi_{f}|\rho|\psi_{f}\rangle \textrm{d}\theta_1$, where the integration is numerically done for 1001 input states with $\theta_1$ being uniformly distributed within $[0, 2\pi]$, and $\rho$ is a numerically simulated density matrix of the qubit system. In our simulation, we select the decay and dephasing rates in the range of $\kappa=\kappa_-=\kappa_z\in[0,8]\times10^{-4}\Omega$; the maximum value is still well within that reached for current state-of-the-art experiments. As shown in Figs. \ref{composite2}(b)-\ref{composite2}(e), by considering the competition of  decoherence and qubit-frequency-drift-induced error, we find the composite geometric gate with $N=2$ is the best choice for the implementation of the noise-resilient geometric gates in our case. A similar discussion is also valid for the two-qubit case. However, due to the increase of the number of the involved physical qubits, the decoherence effect is more destructive to the gate performance; thus the two-qubit gate in a single-loop way in configuration B will be a better choice and it is good enough. These results will be demonstrated in the physical implementation section.

\section{Physical Implementation}

\begin{figure}[tbp]
  \centering
  \includegraphics[width=\linewidth]{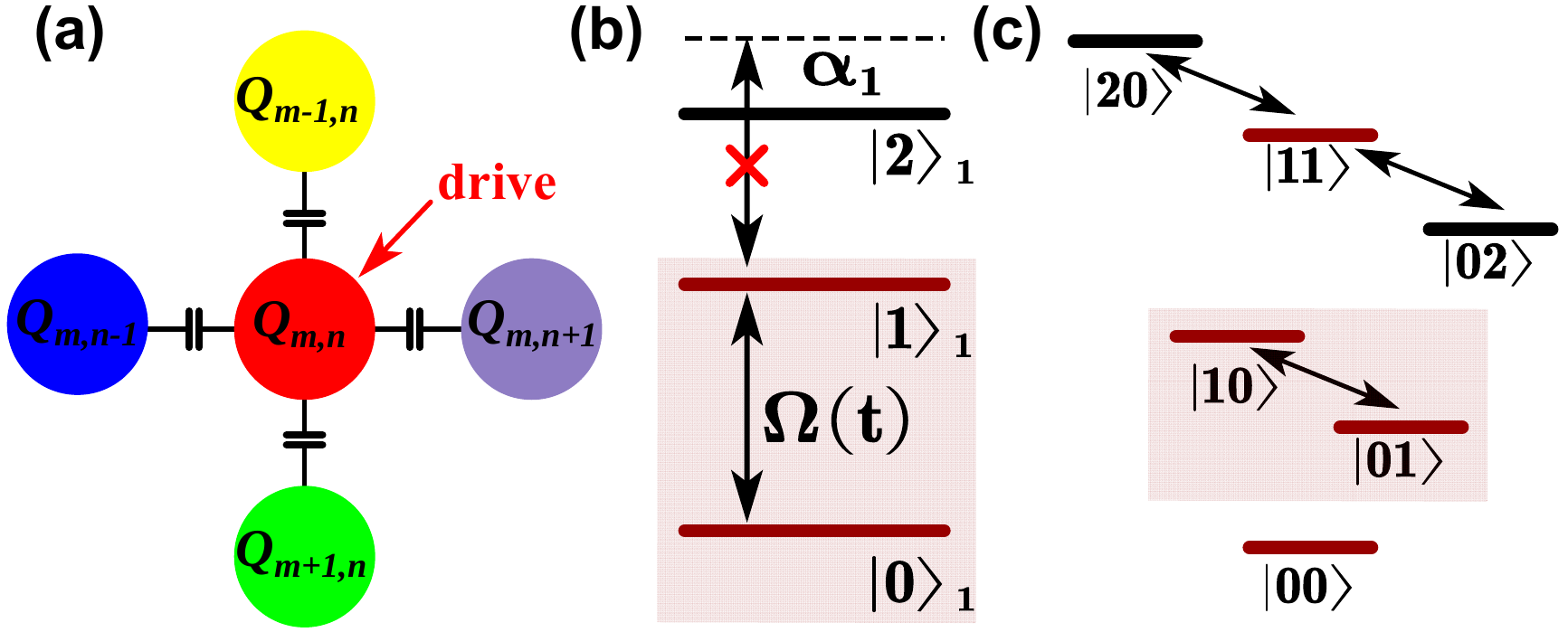}
  \caption{(a) Schematic diagram of scalable 2D square lattice composed of capacitively coupled transmon, where the circles with different colors denote the transmons with different frequency. The subscripts $m$ and $n$ of $Q_{m,n}$ represent the position of the row and column of the transmon qubit on a 2D square lattice, with $m, n \in [1, +\infty]$, respectively. The energy spectrum structures of (b) a driven superconducting transmon with the weak anharmonicity and (c) two parametrically tunable coupled transmons.}
  \label{SQUID}
\end{figure}

\begin{figure*}[tbp]
  \centering
  \includegraphics[width=0.8\linewidth]{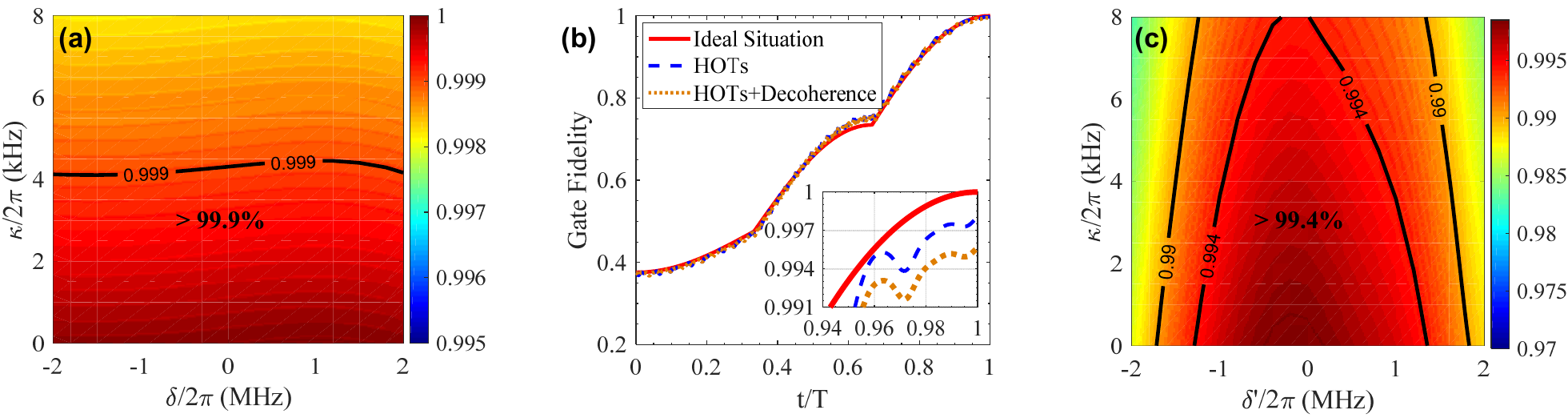}
  \caption{Physical implementation of single- and two-qubit geometric gates. (a) Gate fidelity under the dual influence of the decoherence and the qubit-frequency-drift induced error for the composite geometric NOT gate with $N=2$. (b) Gate-fidelity dynamics with and without considering the effects of the decoherence and high-order oscillating terms (HOTs) for the geometric iSWAP gate. (c) Gate fidelity under the dual influence of the decoherence and the qubit-frequency-drift induced error for the geometric iSWAP gate.} \label{ST}
\end{figure*}

In this section, we will demonstrate the distinct practical merit of our noise-resilient geometric gates on a superconducting circuit, consisting of capacitively coupled transmon qubits as shown in Fig. \ref{SQUID}(a). The superconducting circuit of a transmon qubit is composed of a capacitance and two Josephson junctions, and its energy level structure is depicted in Fig. \ref{SQUID}(b), where the two lowest levels are used as our qubit states. Meanwhile, we also apply the current experimentally mature ``derivative removal via adiabatic gate'' (DRAG) technology \cite{Motzoi2009, DRAGexperiment} to greatly suppress the qubit-leakage error caused by the synchronous coupling of the high-energy levels driven by the external microwave field, thereby implementing independent control of the qubit states $|0\rangle$ and $|1\rangle$. For a driven transmon, its full Hamiltonian can be written as
\begin{eqnarray}
\mathcal{H}_1(t)&=&\sum_{n=1}^{+\infty}\frac {1}{2}\{[2n\omega_1-n(n-1)\alpha_1]|n\rangle_1\langle n| \notag \\
&+& [\sqrt{n}\Omega(t) |n-1\rangle_1\langle n|e^{\textrm{i}(\omega_d t-\phi(t))} + \textrm{H.c.}]\},
\end{eqnarray}
where $\omega_1$ and $\alpha_1$ are the transition frequency and anharmonicity of the first transmon qubit; $\omega_d$ and $\phi(t)$ are the driving frequency and adjustable phase of the microwave field, respectively. In addition, $\Omega(t)=\Omega_{_{\textrm{UD}}}(t)-\textrm{i}\dot{\Omega}_{_{\textrm{UD}}}(t)/2\alpha_1$ is the pulse shape that has been corrected by DRAG, with $\Omega_{_{\textrm{UD}}}(t)=\Omega_0\sin^2(\pi t/\tau)$ being the original uncorrected pulse shape, where time $t \in[0, \tau]$ with $\tau$ being the gate duration.

Moving into the rotating frame with respect to the driving frequency $\omega_d$, and by taking $\omega_d=\omega$, under the rotating-wave approximation, we can obtain an effective interaction Hamiltonian, which is the same as the resonant form of $\mathcal{H}(t)$ in Eq. (\ref{h0}). To check the validation of our scheme under realistic conditions, we next take the decoherence effect and the high-order oscillating terms into consideration to quantitatively analyze how they influence an intended gate operation. We simulate the evolution process numerically by the master equation similar to Eq. (\ref{simulate2}):
\begin{eqnarray}
\label{simulate3}
\dot\rho=-i[\mathcal{H}_1(t), \rho]+\left[{\kappa_-}\mathcal{L}(\lambda_-)+{\kappa_z}\mathcal{L}(\lambda_z)\right]/2,\quad
\end{eqnarray}
where $\lambda_-=\Sigma_{n=1}^{+\infty}\sqrt{n}|n-1\rangle_1\langle n|$ and $\lambda_z=\Sigma_{n=1}^{+\infty}n|n\rangle_1\langle n|$ are the standard lower operator and the projector for the $n$th level, respectively. In our simulation, we choose the geometric composite NOT gate with $N=2$ as a typical example, which corresponds to $\chi=\gamma=\pi/2$ and $\xi_1=0$ in Eq. (\ref{compU}). Considering the current state of the art of experiments, we select the decay and dephasing rates of the transmon in the range of $\kappa=\kappa_-=\kappa_z\in2\pi\times[0,8]$ kHz, and the qubit-frequency-drift-induced error in the range of $\delta\in2\pi\times[-2,2]$ MHz. When the anharmonicity of the transmon and effective coupling strength are $\alpha_1=2\pi \times 320$ MHz and $\Omega_0/2=2\pi \times 30$ MHz, the resulting gate fidelity can exceed $99.90\%$ in the area below the black line, as shown in Fig. \ref{ST}(a). Therefore, our composite scheme can also exhibit the strong noise-resilient feature under realistic conditions, while maintaining high gate fidelity.

Now, we move to the construction of a nontrivial two-qubit geometric gate. Here, we consider a 2D square superconducting qubit lattice, where all the adjacent qubits are capacitively coupled, as shown in Fig. \ref{SQUID}(a). We can choose a pair of the adjacent qubits, such as $Q_{m,n}$ and $Q_{m,n\pm1}$ in the same row (or $Q_{m,n}$ and $Q_{m\pm1,n}$ in the same column), hereafter simplified as $Q_i$ with $i=1, 2$. The form of their capacitive coupling interaction can be expressed as
\begin{eqnarray}\label{h12}
\mathcal{H}_2=&&\sum_{i=1}^2 \sum_{n=1}^\infty \frac{1}{2}\left[2n\omega_i-n(n-1)\alpha_i\right]|n\rangle_i\langle n| \notag \\
          &&+ g_{12}\big(S_1S_2^\dag + S_1^\dag S_2\big),
\end{eqnarray}
where $\omega_i$ and $\alpha_i$ are the associated transition frequency and the intrinsic anharmonicity of transmon $Q_i$, respectively; $g_{12}$ is the coupling strength between transmons $Q_1$ and $Q_2$, and $S_i=\Sigma_{n=1}^{+\infty} \sqrt n|n-1\rangle_i\langle n|$ is the standard lower operator for transmon $Q_i$ of $n$th level. However, due to the effect of the intrinsic weak anharmonicity of the transmon, we need to take the leakage interactions of the high energy level into account, as shown in Fig. \ref{SQUID}(c).

For the implementation of the parametrically tunable coupling between the two adjacent transmons, we drive $Q_1$ by a well-controlled microwave as $\omega_1(t)=\omega_1+\varepsilon\sin(\nu t+\varphi)$. This modulation can be experimentally achieved by biasing the transmon with an ac magnetic flux in a particular dc bias working point \cite{Reagor2018a, Caldwell2018}. To see this, we move the system Hamiltonian of Eq. (\ref{h12}) under the qubit-frequency driving to the interaction picture, where the transformation matrix is defined by $U_t=U_aU_b$ with
\begin{eqnarray}
\label{Ut}
U_a&=&\exp\left[-\textrm{i}\sum_{i=1}^2 \sum_{n=1}^\infty \frac{1}{2}\left[2n\omega_i-n(n-1)\alpha_i\right]|n\rangle_i\langle n| t\right], \notag \\
U_b&=&\exp\left[\textrm{i}\sum_{n=1}^\infty n\beta\cos(\nu t+\varphi)|n\rangle_1\langle n|\right],
\end{eqnarray}
and apply the Jacobi-Anger identity of
\begin{eqnarray}
\exp[-\textrm{i}\beta\cos(\nu t\!+\!\varphi)]\!=\!\!\!\sum_{m=-\infty}^\infty \!\!\!(-\textrm{i})^mJ_m(\beta)\exp[-\textrm{i}m(\nu t\!+\!\varphi)], \nonumber
\end{eqnarray}
in which $\beta\!=\!\varepsilon/\nu$, $J_{-m}(\beta)\!=\!(-1)^m J_m(\beta)$, with $J_m(\beta)$ being Bessel functions of the first kind. The transformed Hamiltonian under driving is
\begin{eqnarray}\label{eq}
\mathcal{H}_{\textrm{int}}\!=\!\!\!&&\sum_{m=-\infty}^\infty \!\!\!J_m(\beta)g_{12} \left\{|10\rangle \langle 01|e^{\textrm{i}\Delta_1t}\!+\!\sqrt 2|11\rangle \langle 02|e^{\textrm{i}(\Delta_1+\alpha_2)t}\right. \nonumber \\
&& \left.+\sqrt 2|20\rangle\langle 11|e^{\textrm{i}(\Delta_1-\!\alpha_1)t} \right\} e^{-\textrm{i}m(\nu t+\varphi+\frac{\pi}{2})}+ \textrm{H.c.},
\end{eqnarray}
where $|mn\rangle=|m\rangle_1\otimes|n\rangle_2$ and $\Delta_1=\omega_1-\omega_2$ is the qubit-frequency difference between the two adjacent transmons $Q_1$ and $Q_2$. It is easy to see that the resonant interaction can be achieved in both the single- or two-excitation subspaces by a different choice of the driving frequency $\nu$. Modulating the qubit-driving frequency to meet $\Delta_1=\nu$, and neglecting the high-order oscillating terms by rotating-wave approximation, then we can obtain the effective Hamiltonian as
\begin{eqnarray}\label{heff1}
\mathcal{H}_{\text{eff}}=g_{\textrm{eff}}|10\rangle\langle 01|e^{-\textrm{i}(\varphi+\frac \pi 2)}+ \textrm{H.c.},
\end{eqnarray}
where the effective coupling strength $g_{\textrm{eff}}=J_1(\beta_1)g_{12}$, which is also the same as the resonant form of $\mathcal{H}(t)$ in Eq. (\ref{h0}). Note that, this effective interaction is within the single-excitation subspace of the coupled qubit system, where quantum information leakage errors are well suppressed. Thus, we can implement the geometric cyclical evolution under configuration B in this effective two-level structure, and the resulting geometric operator remains the same as that in Eq. (\ref{Utau}) within the two-qubit subspace $\{|10\rangle, |01\rangle\}$. By taking $\chi=\gamma=\pi/2$ with $\xi_1=\pi$, the geometric iSWAP gate can be realized, which is a nontrivial two-qubit gate for quantum computation. We further simulate the performance of this gate numerically by the master equation in Eq. (\ref{simulate3}) for the two-qubit case, including the effects of both decoherence and high-order oscillating terms, under the current experimental parameters $\kappa=2\pi\times4$ kHz, the anharmonicity $\alpha_1=2\pi \times 320$ MHz, $\alpha_2=2\pi \times 300$ MHz, coupling strength $g_{12}=2\pi \times 10$ MHz, and the frequency difference of the two qubits being $\Delta_1=2\pi \times 500$ MHz. By these settings, the fidelity of the iSWAP gate can reach $99.56\%$, as shown in Fig. \ref{ST}(b).

In addition, by analyzing the gate fidelity with and without considering the effects of the decoherence and high-order oscillating terms, as shown in Fig. \ref{ST}(b), we can find that the  high-order oscillating terms induce about $0.20\%$ gate infidelity. And the residual infidelity comes from the decoherence of the qubit system. However, for the composite two-qubit gate, although one can still obtain the improvement of the gate robustness, the diploid decoherence and high-order oscillating terms are more destructive, resulting in a gate fidelity below $99.10\%$ for $N=2$; thus the integration of the composite scheme and geometric gate in the two-qubit case is not preferable when pursuing high fidelity. Nevertheless, under the effects of both decoherence and the qubit-frequency-drift-induced error in the form of $\Delta_1+\delta'$, the gate fidelity of our two-qubit geometric gate can still exceed $99.40\%$ in a large parametric range, as shown in Fig. \ref{ST}(c), and it also has a gate-robustness advantage over the corresponding dynamical one (see Appendix A for details). Therefore, our two-qubit geometric gate performance in a single-loop way with configuration B is good enough, where it can also obtain high fidelity and exhibit the strong noise-resilient feature under the current experimental conditions.

\section{Conclusion}


In conclusion, we have proposed a noise-resilient NGQC scheme against the qubit-frequency drifts with a simple setup. We here have applied the path-design strategy to explain in detail why there are two path configurations that can both realize universal geometric gates in a single-loop way. In addition, we can purposefully select the most robust path configuration, and then further integrate with the composite scheme to implement high-fidelity and noise-resilient geometric quantum gates, which are even superior to conventional dynamical gates in terms of  gate robustness. Requiring only the current level of technique, our scheme can be immediately tested in experiments and therefore provides a promising method to achieve high-fidelity geometric manipulation for robust and scalable fault-tolerant quantum computation.

\section*{Acknowledgments}

This work was supported by the Key-Area Research and Development Program of GuangDong Province (No. 2018B030326001), the National Natural Science Foundation of China (No. 11874156), the National Key R\&D Program of China (No. 2016 YFA0301803), Science and Technology Program of Guangzhou (No. 2019050001), the Anhui Provincial Natural Science Foundation (No. 2008085MA20), the Discipline Top-Notch Talents Foundation of Colleges and Universities of Anhui (No. gxbjZD53), and the Research Foundation for Advanced Talents of WXC (No. WGKQ2021004).

\appendix

\begin{figure}[tbp]
  \centering
  \includegraphics[width=0.7\linewidth]{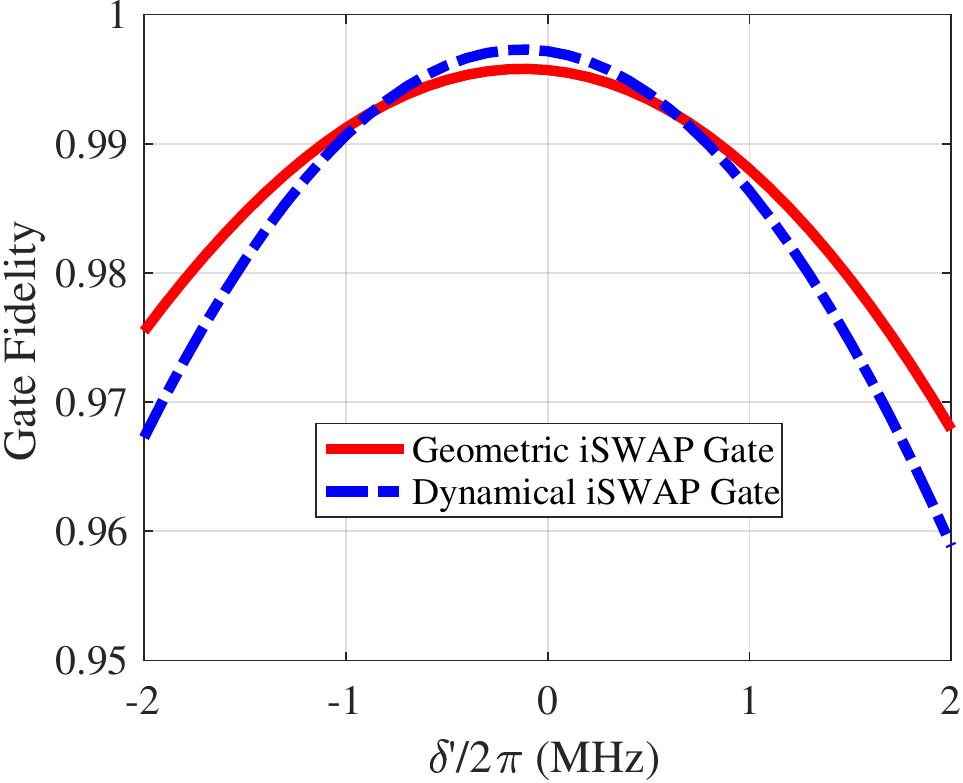}
  \caption{Gate fidelities versus the qubit-frequency-drift-induced $\sigma_z$ error for dynamical and geometric iSWAP gates under the same qubit parameters.} \label{RobustCom}
\end{figure}

\begin{figure}[tbp]
  \centering
  \includegraphics[width=\linewidth]{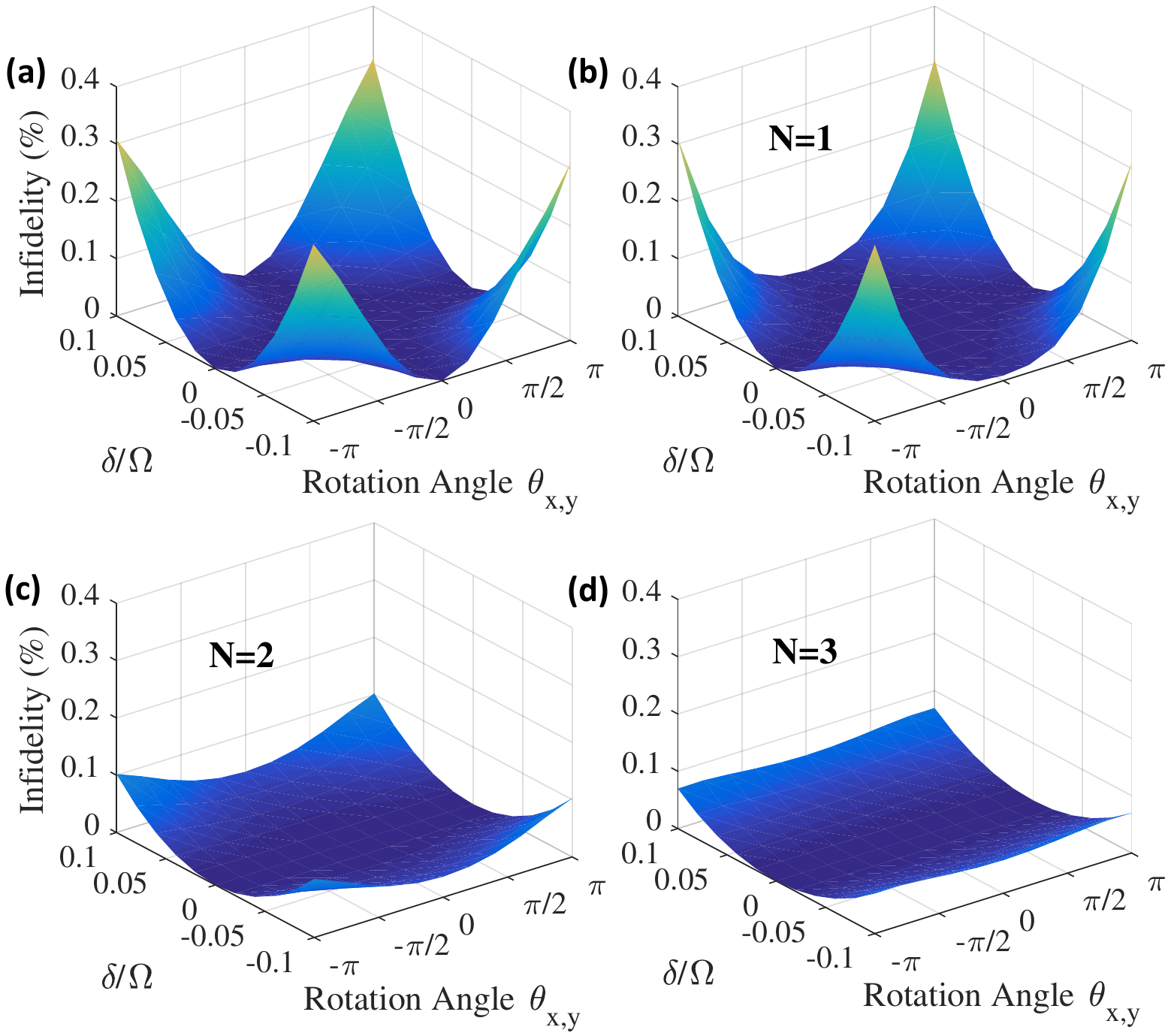}
  \caption{Gate infidelities versus a fluctuant from of qubit-frequency-drift-induced $\sigma_z$ error for (a) conventional dynamical gates, and (b) single-loop $(N\!=\!1)$ and (c,d) composite $(N\!>\!1)$ geometric gates.} \label{composite3}
\end{figure}

\section{Construction of dynamical gates}

For the construction of dynamical gates, we also start from a two-level system driven by a resonant microwave field with $\Delta(t)\!=\!0$; the corresponding Hamiltonian in Eq. (\ref{h0}) is given by $\mathcal{H}(t)=\frac {1} {2}\Omega(t)e^{-\textrm{i}\phi(t)}|0\rangle\langle 1|+\textrm{H.c.}$ Different from the geometric case, the construction of dynamical gates only needs to ensure that the relative phase $\phi(t)$ is constant, i.e., $\phi(t)\!=\!\phi_d$, so that there is no accumulated geometric phase. Thus, the final dynamical evolution operator is
\begin{eqnarray}
\label{EqUd}
U_d(\theta_d,\phi_d)=\left(
\begin{array}{cccc}
 \cos \theta_d         & -\textrm{i}\sin \theta_d e^{-\textrm{i}\phi_d} \\
 -\textrm{i}\sin \theta_d e^{\textrm{i}\phi_d} & \cos \theta_d
\end{array}
\right),
\end{eqnarray}
where $\theta_d=\frac {1} {2}\int^\tau_0 \Omega(t)\textrm{d}t$. In this way, arbitrary X- and Y-axis rotation operations $R^d_{x,y}(\theta_{x,y})$ can also be obtained by setting $\phi_d\!=\!0, \pi/2$ with $\theta_d\!=\!\theta_{x,y}/2$. Note that, to ensure the fairness of our gate robustness comparison, we set the pulse shape and error form to be the same in the cases of both dynamical and geometric gates.

Similarly, for the two-qubit dynamical case, we start from the two-qubit effective Hamiltonian in Eq. (\ref{heff1}). Then, by setting $g_{\textrm{eff}} \textrm{T}\!=\!\pi/2$ and $\varphi\!=\!\pi/2$, the dynamical iSWAP gate can be realized. We here also set the parameters to be the same as that of our geometric gate, i.e., take the decoherence rate $\kappa=2\pi\times4$ kHz and frequency difference of the two qubits as $\Delta_1=2\pi \times 500$ MHz. Under these current experimental parameters, comparing with the dynamical iSWAP gate, our geometric iSWAP gate still possesses the advantage of resisting the qubit-frequency-drift-induced error, as shown in Fig. \ref{RobustCom}.

\section{Gate-robustness for a fluctuant error}

Here, we also consider the case where the qubit-frequency drift is not static but fluctuates on a time-scale shorter than the gate-operation time. We take the fluctuation in the form of $\delta\sin(\pi t/\tau)$ as an example. As shown in Fig. \ref{composite3}, our composite geometric gates can still exhibit an obvious advantage for gate robustness over the conventional dynamical gates as well as single-loop ($N$=1) geometric gates.

\end{document}